\documentclass[twocolumn,aps,floats,superscriptaddress]{revtex4}
\usepackage{amsmath}

\usepackage{epsfig}
\usepackage{graphicx}% Include figure files
\usepackage{dcolumn}% Align table columns on decimal point
\usepackage{bm}% bold math
\renewcommand{\dag}{^{\dagger}}

\def\gapp{\lower.35em\hbox{$\stackrel{\textstyle>}{\sim}$}}
\def\lapp{\lower.35em\hbox{$\stackrel{\textstyle<}{\sim}$}}

\begin{document}
\bibliographystyle{apsrev}
%

%\draft
\title{
Disorder and interaction effects in two dimensional graphene sheets}
\author{T. Stauber$^1$, F. Guinea$^1$, and M.A.H. Vozmediano$^*$}
\affiliation{Instituto de Ciencia de Materiales de Madrid, CSIC, Cantoblanco, E-28049 Madrid, Spain.\\
$^*$Departamento de Matem\'aticas, Universidad Carlos III de Madrid, E-28911 Legan\'es, Madrid, Spain. }
\date{October 31, 2003}
%%%%%%%%%%%%%%%%%%%%%%%%%%%%%%%%%%%%%%%%%%%%%%%%%%%%%%%%%%%%%%%%%%%%%%%%%%%%%
\begin{abstract}
The interplay between different types of disorder and
electron-electron interactions in graphene planes is studied by
means of Renormalization Group techniques. The low temperature
properties of the system are determined by fixed points where the
strength of the interactions remains finite, as in one dimensional
Luttinger liquids. These fixed points can be either stable
(attractive), when the disorder is associated to topological
defects in the lattice or to a random mass term, or unstable
(repulsive) when the disorder is induced by impurities outside the
graphene planes. In addition, we analyze mid-gap states which can
arise near interfaces or vacancies.
\end{abstract}
%%%%%%%%%%%%%%%%%%%%%%%%%%%%%%%%%%%%%%%%%%%%%%%%%%%%%%%%%%%%%%%%%%%%%%%%%%%%%
%
%
%
%%%%%%%%%%%%%%%%%%%%%%%%%%%%%%%%%%%%%%%%%%%%%%%%%%%%%%%%%%%%%%%%%%%%%%%%%%%%%
%%%%%%%%%%%%%%%%%%%%%%%%%%%%%%%%%%%%%%%%%%%%%%%%%%%%%%%%%%%%%%%%%%%%%%%%%%%%%
%%%%%%%%%%%%%%%%%%%%%%%%%%%%%%%%%%%%%%%%%%%%%%%%%%%%%%%%%%%%%%%%%%%%%%%%%%%%%
%%%%%%%%%%%%%%%%%%%%%%%%%%%%%%%%%%%%%%%%%%%%%%%%%%%%%%%%%%%%%%%%%%%%%%%%%%%%%
%
%
\maketitle {\it Introduction.} Graphite is a widely
studied material, which has attracted recent interest due to
the observation of anomalous properties, such as magnetism
or insulating behavior in the direction perpendicular
to the planes in different
samples\cite{Ketal00,Eetal02,KEK02,Ketal02,MHM02,Cetal02,Ketal03,Ketal03b}.

The conduction band of graphite is well described by tight
binding models which include only the $\pi$ orbitals
which are perpendicular to the graphite planes at each
C atom\cite{SW58}. If the interplane hopping is
neglected, this model describes a semi metal, with zero
density of states at the Fermi energy, and where the
Fermi surface is reduced to two inequivalent points in the
Brillouin Zone. The states near these Fermi points
can be described by a continuum model which reduces to the
Dirac equation in two dimensions. Due the the
vanishing of the density of states at the Fermi level,
the long range Coulomb interaction is imperfectly
screened. This implies that a standard perturbative
treatment leads to logarithmic divergences, and to non
trivial deviations from Fermi liquid theory\cite{GGV93,Gon94,Gon99}.
In the strong coupling regime, the model
can exhibit a phase transition which leads to a rearrangement
of the charges and spins within the unit cell, and
which is similar to the chiral symmetry breaking transition
found in field theories\cite{Khv01,Khv01b}.

It is known that disorder significantly changes the states
described by the two dimensional Dirac
equation\cite{CMW96,Cetal97,HD02}, and, usually, the density of
states at low energies is increased. Lattice defects, such as
pentagons and heptagons, or dislocations, can be included by means
of a non Abelian gauge field\cite{Gon93,Gon01}. In general,
disorder enhances the effect of the interactions. In addition, a
graphene plane can show states localized at
interfaces\cite{WS00,W01}, which, in the absence of other types of
disorder, lie at the Fermi energy. Changes in the local
coordination can also lead to localized states\cite{OS91}.

The present work attempts to study, within the same footing,
the role of long range interactions and disorder.
This problem has already been studied in relation with
critical points between integer and fractional fillings
in the Quantum Hall Effect\cite{YS98,Y99}, and we will
be able to translate some of the results there to the
problem at hand. We find, as in\cite{Y99} a rich phase
diagram, with different fixed points. The stability of
these fixed points depends on the nature of the disorder.
Finally, we discuss the changes introduced in this
picture by the possible existence of localized states
near the Fermi edge induced by strong distortions of the
lattice.

{\it The model: Coulomb interaction and disorder.} We describe
the electronic states within each graphene plane
by two two-component spinors associated to the two
inequivalent Fermi points in the Brillouin Zone. They are
combined to a four component (Dirac) spinor. These
spinors obey the massless Dirac equation. The Hamiltonian of
the free system is:
\begin{align}
H_0&=iv_F\int d^2x \overline\Psi(\vec x)\vec\gamma \cdot\vec\nabla\Psi(\vec x)
\end{align}
where $\overline\Psi\equiv \Psi\dag\gamma_0$
with the $4\times4$ matrix
$\gamma_0\equiv\sigma_3\otimes\sigma_3$.
We further have
$\vec\gamma\equiv(\gamma_1,\gamma_2)=(-i\sigma_2,i\sigma_1)\otimes\sigma_3$.
The $\sigma_\mu$ denote the usual Pauli matrices such that
$\{\gamma_\mu,\gamma_\nu\}=2g_{\mu,\nu}{\bf{1}}_{4\times4}$,
$g_{\mu,\nu}$ denoting the Minkowski tensor where $g_{0,0}=1$,
$g_{i,i}=-1$ with $i=1,2$, and zero otherwise.

The long range Coulomb interaction in terms of the Dirac spinors reads
\begin{align}
H_{ee}&=\frac{v_F}{4\pi}\int d^2x d^2x'\overline\Psi(\vec x)\gamma_0\Psi(\vec x)\frac{g}{|\vec x-\vec x'|}
\overline\Psi(\vec x')\gamma_0\Psi(\vec x')
\end{align}
where $g=e^2/v_F$ is the dimensionless coupling constant.

In order to describe disorder effects, the Dirac spinors are
coupled to a gauge field $A(\vec x)$,
\begin{align}
\label{Hdisorder}
H_{disorder}=\frac{v_\Gamma}{4}\int d^2x
\overline\Psi(\vec x)\Gamma\Psi(\vec x)A(\vec x)
\end{align}
where $v_\Gamma$ characterizes the strength and the
$4\times4$ matrix $\Gamma$ the type of the vertex.
In general, $A ( \vec x )$ is a quenched, Gaussian variable with the
dimensionless variance $\Delta$, i.e.,
\begin{align}
\label{Gauss}
\langle A(\vec x)\rangle=0\quad,\quad\langle
A(\vec x)A(\vec x')\rangle=\Delta\delta^2(\vec x-\vec x')\quad.
\end{align}
We will discuss five different types of disorder
which are associated to the five mutually
anticommuting $4\times4$ matrices plus the unity matrix:
i) For a random chemical potential, the $4\times4$
matrix $\Gamma$ is given by $\Gamma=\gamma_0$.
The long range components of this
 type of disorder do not induce transitions between the
two inequivalent Fermi points.
This type of disorder will yield an unstable fixed line.
ii) A random gauge potential involves the $4\times4$ matrices
$\Gamma=i\gamma_1$ and $\Gamma=i\gamma_2$.
This type of disorder will yield a stable fixed line which
is linear in the $(g,\Delta)$-plane.
iii) (a) A fluctuating mass term is described by
$\Gamma={\bf{1}}_{4\times4}$.
(b) Topological disorder is given by $\Gamma=i\gamma_5$ with
$\gamma_5={\bf{1}}_{2\times2}\otimes\sigma_2$.
This type of disorder is associated to the existence
of pentagons and heptagons,
or, more generally, to local distortions of the lattice
axes\cite{Gon93,Gon01}.
This term can thus be represented by a gauge potential which induces
transitions between the two Fermi points.
(c) To complete the discussion, we also mention
$\Gamma=i\tilde\gamma_5$
where $\tilde\gamma_5={\bf{1}}_{2\times2}\otimes\sigma_1$.
This vertex type
can be related to an imaginary mass that couples the
two inequivalent Fermi points.
All these types of disorder will yield a stable
fixed line which is cubic in the $(g,\Delta)$-plane.

{\it Renormalization of the effective couplings.}
To discuss the renormalizability of the theory, we will
first treat the disorder gauge-field as an external
potential and do not consider the average over different
realizations of this field. The free, massless Dirac
propagator is given by\cite{Gon94}:
\begin{align}
G_0(\omega,\vec p)&=-i\int \frac{d\omega}{2\pi}\frac{d^2p}{(2\pi)^2}
e^{i\omega t-i\vec p\vec x}
\langle T\Psi(t,\vec x)\overline\Psi(0,0)\rangle\notag\\
&=\frac{-1}{\gamma_0\omega-v_F \vec \gamma\cdot\vec p+i0}\quad.
\end{align}

\begin{figure}[t]
  \begin{center}
       \epsfig{file=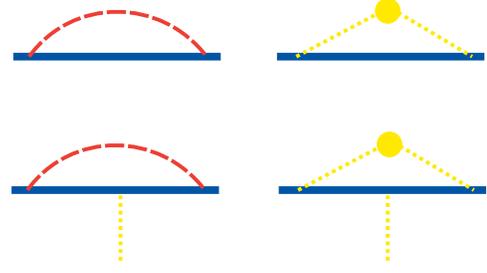,height=3.5cm}
    \caption{Top: Self energy corrections. Dashed line, Coulomb interaction. Dotted line,
ph    external disorder. Bottom: Renormalization of the vertex for the external disorder.}
    \label{Oneloop}
\end{center}
\end{figure}

Within one loop and without averaging over the disorder potential,
only two diagrams need to be considered, i.e., the self-energy of
the fermion propagator due to electron-electron interaction and
the vertex correction of the ``external'' gauge field that couples
to the Dirac bilinear $\overline\Psi\Gamma\Psi$. Both diagrams are
shown on the left hand side of Fig. [\ref{Oneloop}].

The Fermi velocity is renormalized by the top left diagram of Fig.
[\ref{Oneloop}] as $v_F=Z_{v_F}\tilde v_F$
with $Z_{v_F}=1-g/(16\pi\varepsilon)$, using dimensional
regularization with $\varepsilon\to0$\cite{Gon94}.
Notice that the vertex of the Coulomb interaction is
not renormalized within one-loop order.

Whether or not $v_\Gamma$ needs to be  renormalized
depends on the type of the disorder: i) For a random
chemical potential with $\Gamma=\gamma_0$, the right
diagram at the top of Fig. [\ref{Oneloop}] is
non-divergent. This statement is equivalent to the one
that there is no vertex correction of the Coulomb
interaction. We thus set $v_\Gamma=v_1$ with the
flow-invariant velocity $v_1$. ii) For a random
gauge potential
($\Gamma=i\gamma_1,i\gamma_2$), the vertex is renormalized
by the same factor as the Fermi velocity, i.e.,
$Z_\Gamma=1-g/(16\pi\varepsilon)$. This fact is related to
the conservation of the current. We can thus set
$v_{\Gamma}=v_F$. iii) For a random mass term
$\Gamma={\bf{1}}_{4\times4}$, topological disorder
$\Gamma=i\gamma_5$, and $\Gamma=i\tilde\gamma_5$,
the vertex strength $v_\Gamma$ is renormalized by
$Z_\Gamma=1-g/(8\pi\varepsilon)$ which follows
from i) and ii). Within one loop-level, we can thus set
$v_\Gamma=v_F^2/v_3$ with the flow-invariant velocity $v_3$.

Notice that - irrespective of the type of disorder -
our model depends on a single
 renormalized parameter, $v_F$.

{\it Averaging over disorder.} As it was discussed in Ref. \onlinecite{Gon94},
there is no wave function renormalization $Z_\Psi$ to leading order in $g$.
Including self energy corrections due to averaging over an
ensemble of various realizations of the gauge field $A(\vec x)$, the wave function gets
renormalized\cite{Gon01}. The self energy due to disorder is shown at the
top right of Fig. [\ref{Oneloop}] and reads
\begin{align}
\Sigma_{\Gamma}(\omega)=\Delta \frac{v_\Gamma^2}{16}
\int\frac{d^2p}{(2\pi)^2}\Gamma G_0(\omega,\vec p)\Gamma\quad.
\end{align}
The wave function renormalization is independent of the
vertex type and yields $Z_\Psi=1-\Delta v_\Gamma^2/(32\pi v_F^2\varepsilon)$,
where again we use dimensional regularization with $\varepsilon\to0$.

\begin{figure}
%  \begin{center}

       \epsfig{file=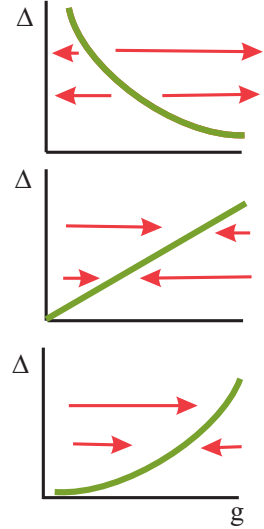,height=7cm}
    \caption{One-loop phase diagram for two-dimensional massless Dirac spinors
including long-ranged electron-electron interaction $g$ and
disorder $\Delta$. Top: Random chemical potential
($\Gamma=\gamma_0$). Center: Random gauge potential
($\Gamma=i\gamma_1,i\gamma_2$). Bottom: Random mass term
($\Gamma={\bf{1}}_{4\times4}$), topological disorder
($\Gamma=i\gamma_5$), and $\Gamma=i\tilde\gamma_5$.}
    \label{Phasediagram}
%\end{center}
\end{figure}

The wave function renormalization $Z_\Psi$ also changes the
renormalization factor of the Fermi velocity as we keep
$G_{0,Z}^{-1}-\Sigma$ invariant with
\begin{align}
G_{0,Z}^{-1}(\omega,\vec p)\equiv-Z_\Psi
(\gamma_0\omega-Z_{v_F}v_F\vec\gamma\cdot\vec p)\quad.
\end{align}

We also need to discuss the vertex corrections due to averaging
over the disorder. The diagram which renormalizes the vertex of
the external gauge field is shown at the bottom of Fig.
[\ref{Oneloop}]. Also the vertex of the Coulomb interaction is
being renormalized (not shown). The vertex correction to $\Gamma$
due to averaging over the disorder type $\Gamma'$ is given by
\begin{align}
V_\Gamma(\omega,\vec k)=\Delta \frac{v_{\Gamma'}^2}{16}
\int\frac{d^2p}{(2\pi)^2}\Gamma' G_0(\omega,\vec p)\Gamma G_0(\omega,\vec p+\vec k)\Gamma'.
\end{align}
This expression depends on the given combination of vertices, 
but does not result in a renormalization of the electric charge ($\Gamma=\gamma_0$).
In the case, where there is only one type of disorder, the
vertex correction exactly compensates the effect of
$Z_\psi$ on $Z_{v_F}$ such that the relation between $v_\Gamma$ and $v_F$ 
remains valid, i.e., i) $v_\Gamma=v_1$,
ii) $v_{\Gamma}=v_F$, and iii) $v_\Gamma=v_F^2/v_3$
with $v_1,v_3$ constant.

{\it Phase diagrams.}
Disorder thus only changes the flow of the Fermi velocity due to
wave function renormalization, i.e., $Z_{v_F}\to Z_{v_F}/Z_\Psi$.
From the beta-function $\beta_{v_F}=\Lambda\partial_\Lambda
Z_{v_F}\tilde v_F$, we obtain the following flow equation
for the effective Fermi velocity $v_F^{eff}$
($\ell=\ln \Lambda/\Lambda_0\sim1/\varepsilon$):
\begin{align}
\label{FlowLutt}
\frac{d}{d\ell}\frac{v_F^{eff}}{\tilde v_F}=
\frac{1}{16\pi}\left[\frac{e^2}{v_F^{eff}}-
\frac{\Delta}{2}\left(\frac{v_\Gamma^{eff}}{v_F^{eff}}\right)^2\right]
\end{align}

We can now discuss the phase diagram for the various types of disorder:
 i) For a random chemical potential $(\Gamma=\gamma_0)$,
$v_\Gamma=v_1$ remains constant under
renormalization group transformation. There is thus an unstable fixed
line at $v_F^*=v_1^2\Delta/(2e^2)$. In the $(g,\Delta)$-plane,
the strong-coupling and the weak-coupling phases are separated
by a hyperbola, with the critical electron interaction
$g^*=e^2/v_F^*=2e^4/(v_1^2\Delta)$.
ii) A random gauge potential involves the vertices $\Gamma=i\gamma_1,i\gamma_2$.
The vertex strength renormalizes as $v_\Gamma=v_F$.  There is thus an
attractive Luttinger-like fixed point for each disorder correlation strength
$\Delta$ given by $v_F^*=2e^2/\Delta$ or $g^*=\Delta/2$.
iii) For a random mass term $\Gamma={\bf{1}}_{4\times4}$, topological disorder
$\Gamma=i\gamma_5$, and $\Gamma=i\tilde\gamma_5$, we have $v_\Gamma=v_F^2/v_3$.
There is thus again an attractive Luttinger-like fixed point for each
disorder correlation strength $\Delta$ given by $v_F^*=\root 3\of{2v_3^2e^2/\Delta}$
or $g^*=\root3\of{\Delta e^4/(2v_3^2)}$.

To make connection to previous work\cite{Y99}, we define
$\widetilde\Delta\equiv\frac{\Delta}{2}(v_\Gamma^{eff}/v_F^{eff})^2$. This
yields the linear fixed line $g^*=\widetilde\Delta$ with i)
$\widetilde\Delta\propto g^2$, ii)
$\widetilde\Delta\propto\text{const.}$, and iii)
$\widetilde\Delta\propto g^{-2}$. For one disorder type, our
results thus agree with the ones by Ye.

{\it Localized states.} The tight binding model defined by the
$\pi$ orbitals at the lattice sites can have edge states when the
sites at the edge belong all to the same
sublattice\cite{WS00,W01,Yetal03}. These states lie at zero
energy, which, for neutral graphene planes, correspond to the
Fermi energy. The states at zero energy are localized in one of
the two interpenetrating sublattices which can be defined in the
honeycomb structure. In terms of each of the two component
spinors,  in the continuum approximation used here, a possible
state localized in one edge can be written as $\Psi ( {\bf
\vec{r}} ) \equiv A ( e^{i k z} , 0 )$,  where $z = x + i y$,  we
assume that the edge lies at $y = 0$, and that the graphene fills
the upper half plane, $y \ge 0$.

\begin{figure}
    \epsfig{file=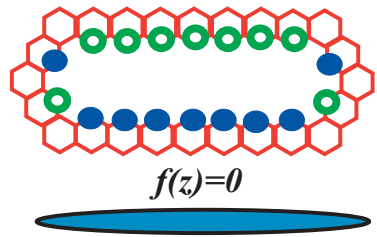
,height=3.5cm}
    \caption{Top: example of a crack in a graphene plane.
The atoms at the upper edge and those at the lower edge belong to
different sublattices. Bottom: approximate cut in the complex plane
which can be used to represent this crack at long distances. See text
for details.}
    \label{graphite_crack}
\end{figure}

In a strongly disordered sample, large defects made up of many
vacancies can exist. These defects give rise to localized states,
when the termination at the edges is locally similar to the
surfaces discussed above. Note that, if the bonds at the edges are
saturated by bonding to other elements, like hydrogen, the states
at these sites are removed from the Fermi energy, but a similar
boundary problem arises for the remaining $\pi$ orbitals. A
particular simple example is given by the crack shown in
Fig.[\ref{graphite_crack}]. The boundary conditions  are such that
$\Psi ( z ) \equiv [ f ( z ) , 0 ]$ where $f ( z ) = 0$ at the
surface of the crack, because it has edges of the two sublattice
types. Possible edge states are: $\Psi_n \equiv \left\{ {\rm Re}
\left[ \frac{A}{z^n \sqrt{z^2 - a^2}} \right] , 0 \right\}$. A
similar solution is obtained by exchanging the upper by the lower
spinor component, and replacing $z \leftrightarrow \bar{z}$.
Because of the discreteness of the lattice, the values of $n$
should be smaller than the number of lattice units spanned by the
crack.

These states are half filled in a neutral graphene plane.
In the absence of electron interactions, this leads
to a large degeneracy in the ground state. A finite local repulsion
will tend to induce a ferromagnetic alignment of the electrons
occupying these states, as in similar cases with degenerate
bands\cite{Vetal99}. Hence, we can assume that the presence of
these states leads to magnetic moments localized near the
defects.

We now have to analyze the influence of
 these magnetic moments in conduction band described in the previous sections.
The hopping between the states involved in the formation of these
moments and the delocalized states in the conduction band vanishes
when the localized states lie at zero energy. Hence, an
antiferromagnetic Kondo coupling will not be induced. The
localized and conduction band states, on the other hand, coexist
at the same lattice sites. The presence of a finite local
repulsion, $U$, will lead to a ferromagnetic coupling. The local
coupling, at site $i$, between the localized states and the
conduction band is proportional to $U \sum_j \rho_{i,j}$, where
$\rho_{i,j}$ is charge of state $j$ at site $i$. The conduction
electrons will mediate an RKKY interaction between the localized
moments:
\begin{equation}
J_{RKKY} ( {\bf \vec{r}} ) \sim
U^2  \int
d^2 {\bf k} e^{i {\bf \vec{k}} {\bf \vec{r}}
} \chi ( {\bf \vec{k}} ) \sim U^2 \frac{a^4}{v_F | {\bf \vec{r}} |^3}
\label{RKKY}
\end{equation}
 Where the static
susceptibility is
 $\chi ( {\bf \vec{k}} ) \propto | {\bf \vec{k}} |$\cite{GGV97},
and $a$ is the lattice constant. It is interesting to note that,
due to the absence of a finite Fermi surface, the RKKY interaction
in eq.(\ref{RKKY}) does not have oscillations. Hence, there are no
competing ferro- and antiferromagnetic couplings, and the magnetic
moments will tend to be ferromagnetically aligned. Asuming $U \sim
\hbar v_F / a \sim 1 - 3$eV, this argument leads to a Curie
temperature $T_C \sim U \langle a^3 / d^3 \rangle$, where $d$ is
the distance between defects.

{\it Conclusions.} In this work, we estimated the effect of
disorder in graphene sheets in the presence of long-ranged
electron-electron interaction. The disorder term can be
characterized by five different vertices which lead - within a
one-loop level - to three distinct phase diagrams. A random
chemical potential divides the phase diagram into a
strong-coupling and weak-coupling regime by an unstable fixed
line. A random gauge potential as well as a random mass term and
topological disorder alter the electronic structure of
two-dimensional graphene sheets as it drives the system towards a
stable, Luttinger-like fixed point. As extensively analyzed
previously (see, for instance\cite{Gon99}) the dimensionless
Coulomb coupling, for graphene planes is $g = e^2 / ( \epsilon_0
v_F ) \sim 1$. Randomness in the chemical potential, due to local
defects, like impurities or vacancies, lead to $\Delta \sim ( V  /
\hbar v_F )^2 c$, where $V$ is the strength of the local potential
and $c$ is the concentration ($\Delta \sim c a^2$ for vacancies,
where $a$ is the lattice constant). Lattice defects, such as
inclinations or disclinations, induce an effective gauge
potential, whose strength, for dislocations, can be expressed as
$\Delta \sim c b^2$\cite{Gon01}, where $b$ is the Burgers vector
of the dislocation. The actual strength of the disorder, however,
can be much larger for experiments performed on irradiated
samples\cite{Hetal03}. Hence, it is likely that graphene sheets
are in the intermediate range coupling, and we can expect a
qualitative behavior like that reported here. Note that the
presence of disorder at intermediate scales will suppress the
chiral symmetry breaking transition expected for pure
graphene\cite{Khv01,Khv01b}.

{\it Acknowledgments.} T.S. is supported by the DAAD-Postdoctoral
program. Funding from MCyT (Spain) through grant
MAT2002-0495-C02-01 is also acknowledged. M.A.H.V. thanks A.
Ludwig and C. Mudry for very useful conversations on disordered
systems. We also thank P. Esquinazi for many illuminating
discussions.
\bibliography{disorder23}
\end{document}